\documentclass[10pt,letterpaper]{article}

\usepackage{cogsci}
\usepackage{pslatex}
\usepackage{apacite}
\usepackage{amsmath, amsfonts, amssymb, bbm}
\usepackage{graphicx, subcaption}

\usepackage{dsfont}
\usepackage{tikz}
\usepackage{tkz-graph}
\usepackage[colorinlistoftodos]{todonotes}
\usepackage[skip=4pt]{caption}
\usepackage{enumitem}
\usepackage{url}
\usepackage{verbatim}

\title{Efficiency of learning vs. processing: Towards a normative theory of multitasking}
 
\author{{\large \bf Yotam Sagiv (ysagiv@princeton.edu), Sebastian Musslick  (musslick@princeton.edu),} \\
  {\large \bf Yael Niv (yael@princeton.edu), Jonathan D. Cohen (jdc@princeton.edu)} \\
  Princeton Neuroscience Institute \\
  Princeton University
}

\begin{document}

\maketitle

\begin{abstract}
A striking limitation of human cognition is our inability to execute some tasks simultaneously. Recent work suggests that such limitations can arise from a fundamental tradeoff in network architectures that is driven by the sharing of representations between tasks: sharing promotes quicker learning, at the expense of interference while multitasking. From this perspective, multitasking failures might reflect a preference for learning efficiency over multitasking capability. We explore this hypothesis by formulating an ideal Bayesian agent that maximizes expected reward by learning either shared or separate representations for a task set. We investigate the agent's behavior and show that over a large space of parameters the agent sacrifices long-run optimality (higher multitasking capacity) for short-term reward (faster learning). Furthermore, we construct a general mathematical framework in which rational choices between learning speed and processing efficiency can be examined for a variety of different task environments.

\end{abstract}

\section{Introduction}
The human brain's ability to simultaneously perform distinct tasks contains a curious tension. On one hand, we are able to concurrently carry out a large number of actions (e.g.\ breathe, speak, chew gum, etc.)\ seemingly without exerting any effort. In contrast, some behaviors defy parallel execution (e.g.\ solving calculus problems and constructing shopping lists) and require serialization to successfully execute. 

The distinction between sets of tasks that can be executed concurrently and those that cannot is often referred to in terms of a fundamental distinction between controlled and automatic processing~\cite{posnerr,shiffrin1977}. Early theories attributed the inability to carry out multiple control-demanding tasks in parallel to reliance on a single, limited-capacity, serial processing mechanism -- a hypothesis that has continued to dominate major theories of cognition \cite<e.g.,>{anderson2013}. The ``multiple-resource hypothesis'' presents a challenge to this view, arguing that multitasking limitations may reflect competition for the use of local resources (e.g., shared task-specific representations) by sets of tasks, rather than common reliance on a central control mechanism \cite{allport1972,feng2014,navon1979economy,meyer1997,musslick2016,salvucci2008threaded,wickens1991processing}. Under this view, the role of cognitive control is to resolve such conflicts when they arise by limiting processing to only a single task at a time \cite{cohen1990,botvinick2001conflict}.  That is, limiting processing is the \emph{purpose of control}, rather than a reflection of a constraint on the control system itself. Recent computational work has provided a formal grounding for this argument, showing that even modest amounts of overlap between task representations can drastically limit the number of tasks a network can engage at the same time without invoking interference among them \cite{feng2014,musslick2016,PetriInPrep}. Critically, this number appears to be relatively insensitive to the size of the network. 

The findings above raise an important question: insofar as shared representation between tasks impose limitations on multitasking, why would a neural system prefer shared representations over separate ones? Insights into this question can be gained from the machine learning literature, where the learning of shared representations between tasks is considered a desirable outcome \cite{baxter1995learning,caruana1998,bengio2013representation}. For instance, work on multi-task learning\footnote{Note that the term 'multi-task' differs from the term 'multitasking'. The former refers to the paradigm of training the same network on multiple tasks, whereas the latter refers to the process of carrying out multiple tasks concurrently.} suggests that shared representations between tasks promote faster learning, as well as better generalization performance across tasks \cite{caruana1997multitask,collobert2008unified}. Moreover, learning dynamics in neural networks themselves promote the learning of shared representation based on shared structure in the task environment \cite{hinton_learning_1986,saxe2013learning,Musslick_et_al_2017}. Thus, there appears to be a fundamental tradeoff in neural networks between the efficiency of learning (and generalization) on the one hand, and the efficiency of processing (i.e.,~multitasking capability) on the other hand \cite{Musslick_et_al_2017}.

The tradeoff between learning and processing efficiency constitutes an optimization problem that is dependent on the demands of the task environment. The work described here examines this optimization problem as a function of critical parameters, such as the differences in rate of learning for shared vs. separated representations, and the benefits gained by parallel over serial task performance.  Analysis of this problem may help provide a formally rigorous, and even normative account of longstanding, well-characterized psychological phenomena, such as the common trajectory in skill acquisition from controlled to automatic processing~\cite{shiffrin1977,logan1980}.

Ideally, our analysis would build on formal characterization of the learning rate for different types of representations, given a specified learning algorithm (e.g.~backpropagation).  However, since this is not immediately available, to construct a probabilistic generative model we begin by assuming simple functional forms for the learning trajectory associated with shared vs.~separated task representations in a multitasking environment, and then use the generative model to define an ideal Bayesian agent that behaves optimally inside that environment. Taken together, the environment and agent models provide a simple, normative framework in which questions about the learning-processing tradeoff can be explored.

\section{A rational model of multitasking}

We begin our analysis of the optimal balance between learning and processing efficiency by formalizing the task environment. We then describe how the agent model chooses between the use of shared vs.~separate representations in that environment to optimize performance, which we define as maximizing reward over the entire horizon of performance.

\subsection{Task Environment}

We consider an environment in which a task can be defined as a process (e.g. naming the color of a stimulus) that maps the dimension of a stimulus (e.g. color) to a particular response dimension (e.g. verbal response). Here we assume that stimuli consist of $N$ dimensions (e.g. color, shape, and texture) and that responses are carried out over $K$ response dimensions (e.g. naming, pointing, or looking), resulting in $NK$ possible tasks in any environment. We adopt a formal definition of multitasking from earlier work \cite{musslick2016,alon2017graph,LesnickFormal}, in which a multitasking condition is defined as the requirement to execute multiple tasks at the same time, none of which share a stimulus or response dimension. Consequently, at most $\min\{N, K\}$ tasks can be carried out concurrently.

The agent is asked to optimize performance over a series of $\tau$ multitasking trials. On each trial, the agent is asked to perform $\alpha$ tasks, where $\alpha$ is drawn from a latent multinomial distribution. We introduce multitasking pressure by specifying a reward schedule that favors concurrent performance of tasks. For every task answered correctly, the agent receives $1$ unit of reward, resulting in $\alpha$ rewards if the agent is able to perform all tasks with maximal accuracy at the same time. However, if the agent chooses instead to perform all tasks sequentially, it loses $jC$ reward units on task $j$, where $j$ indexes the tasks from 0 to $\alpha - 1$ (so that the agent receives $\sum_{j=0}^{\alpha-1} 1-jC$ rewards given maximal accuracy). $C$ is termed the ``serialization cost'' or ``time cost''. We note that this reward schedule is chosen largely for analytical convenience, and is not itself based on a particular normative principle or property of the environment. One alternative could be to set a penalty for serialized execution based on the opportunity cost per time-step. We will extend our results to arbitrary reward schedules in a later section.

Optimization is defined as the choice, on each trial, of a performance strategy that maximizes total future reward; that is, summed over the current trial and the potentially discounted reward anticipated for each future trial. This requires estimating and convolving the expected multitasking requirements over trials, performance for executing the tasks concurrently vs.~individually as a function of the estimated learning rate for each (see below), and the serialization costs associated with performing tasks sequentially.

\subsection{Agent}

\begin{figure}[h]
	\centering
    \includegraphics[scale=0.4]{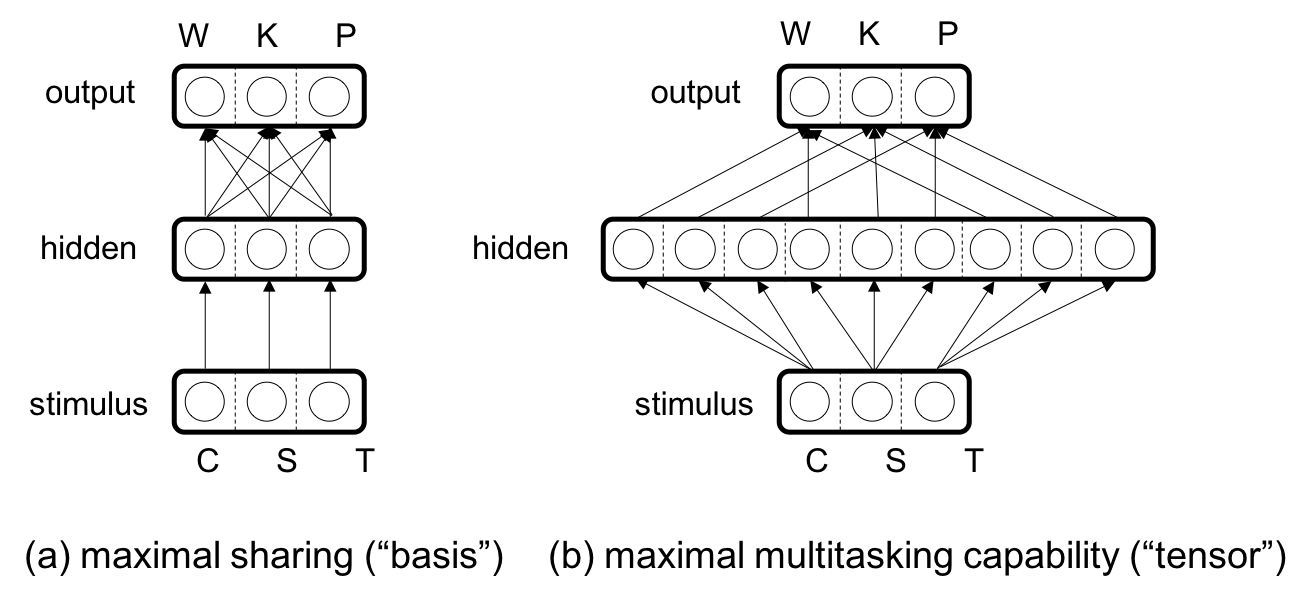}
    \caption{Schematic of network schemes that maximize representation overlap (a) vs. multitasking capability (b). C, S, T designate the stimulus dimensions ("color", "shape", and "texture"), while W, K, P designate the response dimensions ("word", "keyboard", "point"). The hidden-layer representation of the stimulus in (a) is shared for all three tasks involving the same input dimension (minimal basis set representation), whereas in (b) a separate hidden-layer representation is dedicated to each task (tensor product representation).}
    \label{spectrum}
\end{figure}

The agent is considered to be a rational decision-maker that chooses between two independent, trainable processing strategies that result from two extremes of how multiple tasks can be represented in a single network (see Figure~\ref{spectrum}). The first representation strategy is as a minimal basis set, in which all tasks relying on the same stimulus dimension encode the stimuli using the same (shared) set of hidden representations (i.e.~$N$ sets of hidden representations) that are then mapped to the output dimensions for each of the tasks. The second strategy uses tensor product representations, in which each task encodes its stimuli using its own set of (separated) hidden representations (resulting in $NK$ sets of hidden representations) that are mapped to the output dimension for the task. While the minimal basis set provides a more efficient encoding of the stimuli, it does not permit multitasking since the use of shared representations introduces crosstalk between any pair of simultaneously activated tasks~\cite{feng2014,musslick2016,alon2017graph}. Thus, use of the minimal basis set forces a serialization cost of $jC$ reward units for task $j = 1, 2, \dots, \alpha-1$. Conversely, the tensor product representation permits multitasking without interference, since each task is assigned its own set of hidden representations that comprise independent processing pathways in the network. We assume that the agent has the potential to develop both forms of representation, but these must be learned. 

Previous work has shown that, for a set of tasks that are in principle multitaskable, training using shared representations (such as a minimal basis set) leads to faster acquisition than learning separate representations for each task (such as a tensor product), as the former enables the sharing of learning signals across tasks \cite{Musslick_et_al_2017}. We implement these effects by assuming that 1) the agent learns these two types of representations (i.e.~processing strategies) by selecting and executing one or the other on each trial; 2) performance for each strategy improves as a function of the number of trials selected, and 3) learning is faster for the minimal basis set strategy than for the tensor product strategy, as described below. 

To model the learning of tasks, we define a probability of success function (aka ``training function'') for each of the two processing strategies. Let $f_B, f_T : \mathbb{N}_{\geq 0} \to [0, 1]$ denote these training functions for the minimal basis set and tensor product strategies, respectively. These serve as explicit characterizations of the agent's learning dynamics; $f_X(t)$  implements the learning curve by evaluating the probability of success on a given task after representation $X$ has been selected $t$ times. That is, every time the agent chooses to process the tasks in the trial using strategy $X$, the success probability for the task under strategy $X$ increases for the next time-step. More formally, let $x_1, x_2, \dots, x_{t-1}$ be a sequence of $t - 1$ choices of representation. We define the probability that an agent succeeds when employing strategy $X$ on a task on trial $t$ as:
\begin{equation} \label{p_succ_def}
    \mathbb{P}_X(\text{success on a task in trial $t$}) = f_X(\sum_{i = 1}^{t - 1} \mathbbm{1}_{x_i = X}) 
\end{equation}

For convenience, we use the logistic function $f_X(t|~ k, t_0) = \frac{1}{1 + e^{-k(t-t_0)}}$. However, our analysis applies to any learning function that is monotonically increasing and is bounded $0 \leq f_X(t) \leq 1$, for all $t$. As noted above, we assume that learning occurs at a faster rate for the minimal basis set strategy as compared to tensor product strategy, and examine the influence of this difference by exploring a range of values for $k,~t_0$ that together determine the rate of learning.

The agent uses standard Bayesian machinery to infer the expected reward under each representation, and then selects the representation that maximizes total discounted future reward. Specifically, let $\mathbb{E}_X[R]$ denote the expected reward for strategy $X$, $\mathbb{E}_X[R | t]$ denote the expected reward on trial $t$, and $\mu(t)$ be the temporal discounting function. Then we have that $\mathbb{E}_X[R] = \sum_{t=0}^\tau \mu(t) \mathbb{E}_X[R | t]$. Though temporal discounting can be irrational in many contexts, we note that a fully rational agent can be achieved with $\mu(t) = 1$. 

Recall that $\alpha$ is the randomly assigned number of tasks required to be performed on a given trial. By marginalizing over $\alpha$, we get that the expected reward on each individual trial is $\mathbb{E}_X[R|t] = \sum_{i=1}^{\min\{N, K\}} \mathbb{P}(\alpha = i) \mathbb{E}_X[R | t, \alpha = i]$. Thus, the expected rewards for the minimal basis set and tensor product strategies correspond to
\begin{equation} \label{exp_r_trial}
	\begin{aligned}
		\mathbb{E}_B[R|t] &= \sum_{i=1}^{\min\{N, K\}} \mathbb{P}(\alpha = i) \sum_{j=0}^{i-1} \mathbb{P}_B(\text{success}|t)(1 - jC) \\
    	\mathbb{E}_T[R|t] &= \sum_{i=1}^{\min\{N, K\}} \mathbb{P}(\alpha = i) \sum_{j=0}^{i-1} \mathbb{P}_T(\text{success}|t)(1) \\
    \end{aligned}
\end{equation}

In order to compute the expected reward terms in Equation (\ref{exp_r_trial}), the agent must be able to evaluate $\mathbb{P}(\alpha = i)$ and $\mathbb{P}_X(\text{success}|t)$ by inferring the multinomial task distribution, as well as the parameters of each training function $f_X$. The first can be inferred using Bayes' theorem, by keeping track of the number of times each particular $\alpha$ value was seen, in conjunction with a Dirichlet prior (we start from a uniform prior, implying absence of strong {\em a priori} belief about the distribution). 

Inferring the parameters for the two training functions $f_B, f_T$ can similarly be done by tracking the history of successes and failures and then performing a Bayesian logistic regression (intuitively, this can be understood as the agent inferring how fast it will learn). In this model, $k$ and $t_0$ have independent normal priors centered on their true values with high variance. Finally, we assume that the agent already knows $\tau$, the sequential processing cost $~C,$ and the temporal discounting function $\mu(t)$. 

Once the expected values are computed, the agent must select an action. We assume this is done using a standard explore-exploit algorithm, the $\epsilon$-greedy rule, in which the agent picks the action associated with greatest value with probability $1 - \epsilon$, and uniformly otherwise.


\section{Formal analysis of equilibrium}

We begin by analyzing an agent that has perfect knowledge about the task environment and learning rate, in order to assess performance independently of noise that might be generated by an inference process over these factors. This allows us to analytically derive equilibrium conditions under which the agent should be indifferent between the minimal basis set and the tensor product strategies. For this section, we let $N < K$ so that $N = \min\{N, K\}$ without loss of generality.

Observe that the expressions in Equation (\ref{exp_r_trial}) reduce to:
\begin{equation} \label{exp_r_trial_simple}
	\begin{aligned}
		\mathbb{E}_B[R|t] &= f_B(t) \mathbb{E}[g(\alpha, C)] \\
    	\mathbb{E}_T[R|t] &= f_T(t) \mathbb{E}[\alpha]
    \end{aligned}
\end{equation}

\noindent where $g(i, C) = \sum_{j = 0}^{i - 1} (1 - jC)$. Note that $g(i, C)$ encodes the amount of reward accrued by the agent for completing $i$ tasks in a serial fashion with time cost $C$. Plugging Equation (\ref{exp_r_trial_simple}) into the expression for the expected reward of both strategies we can express the condition for which the agent should be indifferent between them:
\begin{equation} \label{equilibrium}
	\frac{\mathbb{E}[\alpha]}{\mathbb{E}[g(\alpha, C)]} = \frac{\sum_{t = 0}^\tau \mu(t)f_{B}(t)}{\sum_{t = 0}^\tau \mu(t)f_{T}(t)}
\end{equation}

An interesting property of this result is that agent-related and environmental parameters are analytically separable. Observe that the expectation terms on the left correspond to the agent's expected reward at asymptotic performance levels, and that the sum terms on the right denote the number of expected successes in a critical time period specified by the conjunction of the temporal discounting function and the training function. The indifference point can be understood intuitively as a surface over which the ratio of expected eventual rewards is equal to the ratio of times at which they are likely to be accrued (discounted by time). That is, the left side contains the ratio of the rewards the agent expects to earn if it is always correct, whereas the right side is a ratio of functions that weight when the agent prefers to receive the rewards.

Recall that $\mathbb{E}[g(\alpha, C)]$ corresponds to $\mathbb{E}[\sum_{j = 0}^{\alpha - 1} (1 - jC)] = \mathbb{E}\Big[\frac{\alpha}{2}\Big(1 + [1 - (\alpha - 1)C]\Big)\Big]$. Since $C$ is a constant, it can be isolated from the expectation in Equation (\ref{equilibrium}) to get an expression for the precise value of the serialization cost that characterizes the indifference surface. That is:
\begin{equation}
    C_{eq} = \frac{2\mathbb{E}[\alpha]\Big(1 - \frac{\sum_{t = 0}^\tau \mu(t)f_T(t)}{\sum_{t = 0}^\tau \mu(t)f_B(t)}\Big)}{\mathbb{E}[\alpha(\alpha-1)]} \label{ceq}
\end{equation}

Equation (\ref{ceq}) provides a rigorous characterization of the tradeoff between basis set and tensor product learning in multitasking environments described in the Introduction:

\begin{enumerate}
    \item As the average number of parallel tasks increases, the cost of serialization must vanish for minimal basis set representations to remain preferable: $\mathbb{E}[\alpha] \to \infty \implies C_{eq} \to 0$. 
     \item As the learning benefit of shared representations diminishes, the value of shared representations disappears. That is, as the ratio between the (discounted) tensor product and basis set training functions approaches unity, for the latter to remain preferable the cost of serialization must tend toward zero: $\frac{\sum_{t = 0}^\tau \mu(t)f_T(t)}{\sum_{t = 0}^\tau \mu(t)f_B(t)} \to 1 \implies C_{eq} \to 0$. 
    \item $\frac{\sum_{t = 0}^\tau \mu(t)f_T(t)}{\sum_{t = 0}^\tau \mu(t)f_B(t)} \to 0 \implies C_{eq} \to \frac{2\mathbb{E}[\alpha]}{\mathbb{E}[\alpha(\alpha - 1)]}$: As the ratio of the discounted training functions for the tensor product and minimal basis set representations approaches 0, the equilibrium-defining serialization cost becomes a function of the number of tasks required to be performed. Particularly, $C_{eq}$ is the serialization cost that sets expected reward for the minimal basis set representation to 0. This implication is not immediately obvious. Consider the task distribution $\mathbb{P}[\alpha = 1] = \mathbb{P}[\alpha = 2] = 1/2$. In this environment, $C_{eq} = 3$ and at asymptotic performance levels, the agent expects to win $1$ reward unit when $\alpha = 1$, or win $-1$ when $\alpha = 2$. This makes sense; if  learning tensor product representations is so much slower than minimal basis set representations that the ratio of the sums goes to 0, the agent is indifferent only if the expected earnings are 0. 
\end{enumerate}

Finally, we note that we have used arbitrary reward functions for the analyses above. However, it is possible to generalize the equilibrium condition in Equation (\ref{equilibrium}) to any stationary reward function (i.e. does not change over the course of the experiment). Let $g_B(j, \theta_B)$ denote any reward function applied independently to each task, with arbitrary dependence on the task's index $j$ and other fixed parameters $\theta_B$. Furthermore, let $G_B(i, \theta_B) = \sum_{j=0}^{i-1} g_B(j, \theta_B)$ be the accumulated reward across a task set consisting of $i$ tasks. Note that previous analysis corresponds to the case $g_B(j, \theta_B) = 1-jC$. Specifically, $g_B$ and $G_B$ are the per-task and cumulative reward functions when the agent executes tasks serially. Finally, define $g_T, G_T$ analogously for the case where the tasks are being processed concurrently. Then a generalized equilibrium condition is: 
\begin{equation}
    \frac{\mathbb{E}[G_T(\alpha, C)]}{\mathbb{E}[G_B(\alpha, C)]} = \frac{\sum_{t = 0}^\tau \mu(t)f_B(t)}{\sum_{t = 0}^\tau \mu(t)f_T(t)} \label{gen_equilibrium} 
\end{equation}

Observe that for $g_B = 1-jC$ and $g_T = 1$, this reduces to the expression in Equation (\ref{equilibrium}). The existence of this generalized equilibrium condition allows a large set of questions to be phrased within this framework. For example, it is easy to include an explicit cost of cognitive control \cite<e.g.,>{shenhav2013, shenhav2017, manohar2015reward} by adding a term to the basis set reward function that implements a cost that increases with the number of tasks executed, or the use of a per-task penalty consisting of the asymptotic-performance opportunity cost (a function exclusively of $\alpha$). 

\section{Numerical analysis with parameter inference}
The analysis above characterized the behavior of an agent with perfect knowledge of the task environment and its learning functions. Here we relax these assumptions, and use numerical simulations\footnote{code available at \url{https://github.com/yotamSagiv/thesis}} to evaluate the behavior of an agent that must infer these parameters. We assess the agent's performance across a series of task environments and learning specifications by crossing a set of reasonable parameter ranges.

We let $\tau = 1000$. We set $C \in [0, 1]$, varying from no punishment to receiving no reward for a correct answer. We use an exponential discounting scheme $\mu(t) = \gamma^{-0.025t}$ for $\gamma \in [0.5, 1.0]$. This covers the range from extreme discounting to no discounting at all. We characterize the training functions as logistic with $f_X(t) = \frac{1}{1 + e^{-0.1(t - t_X)}}$. This allows us to precisely characterize difference in learning rates through the ratio $t_T/t_B$. To that end, we set $t_B = 200$, reflecting the speed of minimal basis set learning, and let $t_T$ vary in $[200, 600]$. We let $N = K = 4$ and define the distribution over tasks as $\mathbb{P}(\alpha = 1) = 0.7,~\mathbb{P}(\alpha = 2) = \mathbb{P}(\alpha = 3) = \mathbb{P}(\alpha = 4) = 0.1$ so that the intensity and frequency of multitasking trials is sufficient to permit either strategy given appropriate parameters. We set $\epsilon = 0.1$ to facilitate early exploration of the tensor product option in the face of immediate rewards due to the minimal basis set option. Finally, we quantify the agent's strategy preference as  $\mathbb{P}(\text{pick }X) = \frac{\text{number of times $X$ was picked}}{\tau}$, and track how $\mathbb{P}(\text{pick basis set})$ varies with the parameters\footnote{We can use Equation (\ref{ceq}) to show that even with weak discounting ($\gamma = 0.90$) and a modest learning rate ratio $t_T/t_B = 2$, the importance of fast training is such that the time cost must nearly equal the reward value ($C_{eq} \approx 0.75$) for indifference in this environment. }. 

\begin{figure}[ht]
	\centering
	\includegraphics[scale=0.23]{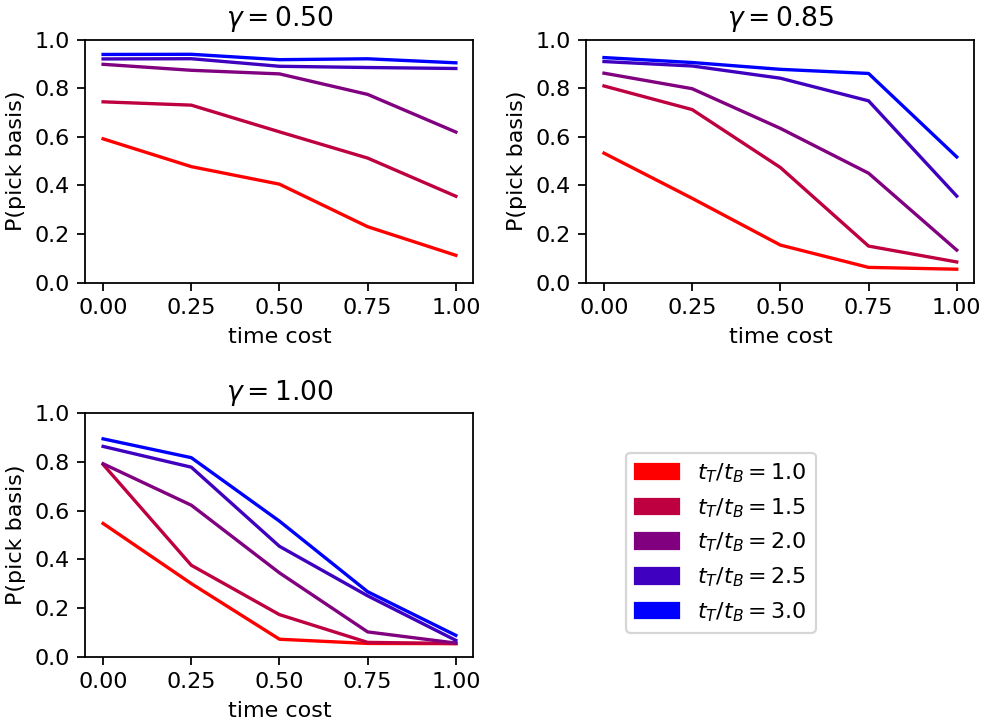}
    \caption{Simulation results for the inference model. $t_T/t_B$ refers to the midpoint ratio of the tensor product and minimal basis set training functions. Time cost denotes the value of $C$. Note that the agent increases their preference for the minimal basis set representation when the time cost is decreased, the learning rate ratio is increased, or gamma is decreased.} 
    \label{data1}
\end{figure}

The results (see Fig.~\ref{data1}) show that there is a broad range of parameters under which the agent will opt for selecting the minimal basis set strategy over the tensor product strategy ($\mathbb{P}(\text{pick basis set}) > 0.5$). These preferences align with the normative analysis of how the parameters should affect overall preference: preference for the minimal basis set strategy increases with relative speed of learning, decreases with serialization cost, and increases with the strength of temporal discounting as indicated by the linear model fit $\mathbb{P}(\text{select basis set}) \sim b_1 \times \frac{t_T}{t_B}+b_2 \times \text{timeCost}+b_3 \times \gamma$ ($b_1=0.25, t(78)=47.26, p<0.001$;~$b_2=-0.52, t(78)=-49.38, p<0.001$;~$b_3=-0.64, t(78)=-35.78, p<0.001$). 

\section{Discussion}

The constraints on human multitasking abilities present an interesting puzzle given the enormous processing capability of the brain. Here, we explored the hypothesis that this reflects a fundamental tradeoff between learning and processing efficiency~\cite{Musslick_et_al_2017}, in which preference for learning to perform a set of tasks faster, which relies on the use of shared representations \cite{caruana1998,baxter1995learning}, comes at the expense of multitasking efficiency~\cite{allport1972,feng2014,navon1979economy,meyer1997,musslick2016,salvucci2008threaded,wickens1991processing}. This tradeoff between the value of shared vs.~separated representations is reminiscent of the complementary learning systems hypothesis~\cite{mcclelland1995}, which proposes the existence of two independent learning mechanisms. The first relies on shared representations to support inference, and the second uses separate representations to avoid the cost of catastrophic interference for memory encoding and retrieval. Thus, the tradeoff between shared and separated representations appears to a fundamental one, that has different consequences in different processing contexts. Here, we have provided a normative analysis of this tradeoff in the context of task performance that, under various assumptions, defines the conditions under which limitations in multitasking ability can be viewed as a result of optimal decision-making. 

Agent behavior in our model was governed by several factors: the distribution of multitasking opportunities within the environment, the cost of serial vs.\ parallel performance, the rate at which each strategy is learned, and the discount rate for future rewards.  The broad range of these factors over which the minimal basis set strategy was optimal suggests that the theory provides a plausible account of why so many skills (e.g. driving a car, playing an instrument) seem to rely on cognitive control and serial execution during acquisition.

Theories of bounded rationality \cite{simon1955behavioral,simon1982models,gigerenzer2008heuristics} assume that suboptimalities in human behavior arise from the use of heuristics rather than full deliberation, given the bounds of limited multitasking capacity and limited available information. Research in artificial intelligence has suggested that such behavior is normative; that is, it may reflect bounded \emph{optimality}, in which an agent maximizes reward per unit time given intrinsic limitations in its computational architecture \cite{russell1995provably}. The principles of bounded optimality are reflected in psychological models of cognition, in which humans perform optimally within the constraints of the cognitive system \cite{griffiths2015rational,gershman2015computational}. Yet, these accounts do not explain why computational limitations exist in the first place, other than the assumption of limited processing power/speed.  The work here suggests that the bounds may arise from a normative response to constraints imposed by tradeoffs intrinsic to \emph{any network architecture}, whether neural or artificial -- specifically, the tradeoff between the advantages of faster learning and generalization provided by shared representations, and the advantages of concurrent parallelism and processing efficiency provided by separated representations \cite{Musslick_et_al_2017}. Under this framework the source of the limitation is not in the brain/computing device, but rather in the fact that time in life is finite (i.e., the benefits of learning a task quickly far outweigh the value of learning it ``optimally''). 

Of course, the model we described is relatively simple, and can be extended in a number of ways. Rather than using a logistic function to characterize learning, it may be more reasonable to scale the benefit of shared representations by the number of tasks \cite<e.g.~as in>{Musslick_et_al_2017}, or to implement the learning dynamics of actual neural networks on similar task spaces. Additionally, a cost of control parameter could be incorporated that scales with the number of tasks being executed and/or the complexity of the task environment~\cite{shenhav2013}. It is also plausible to consider the transfer of learning between the two strategies (i.e. generalization). This may be an important factor in shaping how representations evolve from the minimal basis set to tensor product forms over the course of training, as suggested by some neural evidence~\cite{garner2015}.  

One might also consider meta-learning. The simulated agents learned about their task environment and learning functions, but always began with the same predetermined, static priors. It is possible that repeated experience over different task domains could inform these priors, improving the initial estimates of the learning functions. This would induce a higher rate of convergence to the optimal decision for cases in which the agent's prior experiences are relevant, and might also explain any reluctance to switch away from suboptimal decision-making in contexts where its experience is misleading. Such effects could be informative to similar lines of inquiry regarding separate mechanisms for goal-directed and habitual responding in mammals undergoing instrumental conditioning~\cite{yin2006}. 

In sum, the results presented here strongly support the proposal that constraints in multitasking observed in human performance may arise from a normative approach to an inescapable tradeoff between the value of rapidly acquiring a set of novel skills, and optimizing the efficiency with which these skills can be exercised. Such a normative theory of multitasking may have value not only for understanding human performance, but also for the design of artificial systems. Having a formal language with which to consider the tradeoff between learning efficiency and multitasking capability (and the closely related constructs of controlled vs.~automatic processing) will facilitate precise analysis of the design of autonomous agents that are capable not only of guiding their own actions, but also of learning the best ways of doing so.

\bibliographystyle{apacite}

\setlength{\bibleftmargin}{.125in}
\setlength{\bibindent}{-\bibleftmargin}

\bibliography{CogSci_Template}

\begin{thebibliography}{}

\bibitem [\protect \citeauthoryear {%
Allport%
, Antonis%
\BCBL {}\ \BBA {} Reynolds%
}{%
Allport%
\ \protect \BOthers {.}}{%
{\protect \APACyear {1972}}%
}]{%
allport1972}
\APACinsertmetastar {%
allport1972}%
\begin{APACrefauthors}%
Allport, A.%
, Antonis, B.%
\BCBL {}\ \BBA {} Reynolds, P.%
\end{APACrefauthors}%
\unskip\
\newblock
\APACrefYearMonthDay{1972}{}{}.
\newblock
{\BBOQ}\APACrefatitle {On the division of attention: A disproof of the single
  channel hypothesis} {On the division of attention: A disproof of the single
  channel hypothesis}.{\BBCQ}
\newblock
\APACjournalVolNumPages{Quarterly Journal of Experimental
  Psychology}{24}{2}{225-235}.
\newblock
\begin{APACrefDOI} \doi{10.1080/00335557243000102} \end{APACrefDOI}
\PrintBackRefs{\CurrentBib}

\bibitem [\protect \citeauthoryear {%
Alon%
\ \protect \BOthers {.}}{%
Alon%
\ \protect \BOthers {.}}{%
{\protect \APACyear {2017}}%
}]{%
alon2017graph}
\APACinsertmetastar {%
alon2017graph}%
\begin{APACrefauthors}%
Alon, N.%
, Reichman, D.%
, Shinkar, I.%
, Wagner, T.%
, Musslick, S.%
, Cohen, J\BPBI D.%
\BDBL {}Ozcimder, K.%
\end{APACrefauthors}%
\unskip\
\newblock
\APACrefYearMonthDay{2017}{}{}.
\newblock
{\BBOQ}\APACrefatitle {A graph-theoretic approach to multitasking} {A
  graph-theoretic approach to multitasking}.{\BBCQ}
\newblock
\BIn{} \APACrefbtitle {Advances in Neural Information Processing Systems}
  {Advances in neural information processing systems}\ (\BPGS\ 2097--2106).
\PrintBackRefs{\CurrentBib}

\bibitem [\protect \citeauthoryear {%
Anderson%
}{%
Anderson%
}{%
{\protect \APACyear {2013}}%
}]{%
anderson2013}
\APACinsertmetastar {%
anderson2013}%
\begin{APACrefauthors}%
Anderson, J\BPBI R.%
\end{APACrefauthors}%
\unskip\
\newblock
\APACrefYear{2013}.
\newblock
\APACrefbtitle {The Architecture of Cognition} {The architecture of cognition}.
\newblock
\APACaddressPublisher{}{Psychology Press}.
\PrintBackRefs{\CurrentBib}

\bibitem [\protect \citeauthoryear {%
Baxter%
}{%
Baxter%
}{%
{\protect \APACyear {1995}}%
}]{%
baxter1995learning}
\APACinsertmetastar {%
baxter1995learning}%
\begin{APACrefauthors}%
Baxter, J.%
\end{APACrefauthors}%
\unskip\
\newblock
\APACrefYearMonthDay{1995}{}{}.
\newblock
{\BBOQ}\APACrefatitle {Learning internal representations} {Learning internal
  representations}.{\BBCQ}
\newblock
\BIn{} \APACrefbtitle {Proceedings of the eighth annual conference on
  Computational learning theory} {Proceedings of the eighth annual conference
  on computational learning theory}\ (\BPGS\ 311--320).
\PrintBackRefs{\CurrentBib}

\bibitem [\protect \citeauthoryear {%
Bengio%
, Courville%
\BCBL {}\ \BBA {} Vincent%
}{%
Bengio%
\ \protect \BOthers {.}}{%
{\protect \APACyear {2013}}%
}]{%
bengio2013representation}
\APACinsertmetastar {%
bengio2013representation}%
\begin{APACrefauthors}%
Bengio, Y.%
, Courville, A.%
\BCBL {}\ \BBA {} Vincent, P.%
\end{APACrefauthors}%
\unskip\
\newblock
\APACrefYearMonthDay{2013}{}{}.
\newblock
{\BBOQ}\APACrefatitle {Representation learning: A review and new perspectives}
  {Representation learning: A review and new perspectives}.{\BBCQ}
\newblock
\APACjournalVolNumPages{IEEE Transactions on Pattern Analysis and Machine
  Intelligence}{35}{8}{1798--1828}.
\PrintBackRefs{\CurrentBib}

\bibitem [\protect \citeauthoryear {%
Botvinick%
, Braver%
, Barch%
, Carter%
\BCBL {}\ \BBA {} Cohen%
}{%
Botvinick%
\ \protect \BOthers {.}}{%
{\protect \APACyear {2001}}%
}]{%
botvinick2001conflict}
\APACinsertmetastar {%
botvinick2001conflict}%
\begin{APACrefauthors}%
Botvinick, M\BPBI M.%
, Braver, T\BPBI S.%
, Barch, D\BPBI M.%
, Carter, C\BPBI S.%
\BCBL {}\ \BBA {} Cohen, J\BPBI D.%
\end{APACrefauthors}%
\unskip\
\newblock
\APACrefYearMonthDay{2001}{}{}.
\newblock
{\BBOQ}\APACrefatitle {Conflict monitoring and cognitive control.} {Conflict
  monitoring and cognitive control.}{\BBCQ}
\newblock
\APACjournalVolNumPages{Psychological review}{108}{3}{624}.
\PrintBackRefs{\CurrentBib}

\bibitem [\protect \citeauthoryear {%
Caruana%
}{%
Caruana%
}{%
{\protect \APACyear {1997}}%
}]{%
caruana1997multitask}
\APACinsertmetastar {%
caruana1997multitask}%
\begin{APACrefauthors}%
Caruana, R.%
\end{APACrefauthors}%
\unskip\
\newblock
\APACrefYearMonthDay{1997}{}{}.
\newblock
{\BBOQ}\APACrefatitle {Multitask learning} {Multitask learning}.{\BBCQ}
\newblock
\APACjournalVolNumPages{Machine learning}{28}{1}{41--75}.
\PrintBackRefs{\CurrentBib}

\bibitem [\protect \citeauthoryear {%
Caruana%
}{%
Caruana%
}{%
{\protect \APACyear {1998}}%
}]{%
caruana1998}
\APACinsertmetastar {%
caruana1998}%
\begin{APACrefauthors}%
Caruana, R.%
\end{APACrefauthors}%
\unskip\
\newblock
\APACrefYearMonthDay{1998}{}{}.
\newblock
{\BBOQ}\APACrefatitle {Multitask Learning} {Multitask learning}.{\BBCQ}
\newblock
\BIn{} S.~Thrun\ \BBA {} L.~Pratt\ (\BEDS), \APACrefbtitle {Learning to Learn}
  {Learning to learn}\ (\BPGS\ 95--133).
\newblock
\APACaddressPublisher{Boston, MA}{Springer US}.
\newblock
\begin{APACrefDOI} \doi{10.1007/978-1-4615-5529-2_5} \end{APACrefDOI}
\PrintBackRefs{\CurrentBib}

\bibitem [\protect \citeauthoryear {%
Cohen%
, Dunbar%
\BCBL {}\ \BBA {} McClelland%
}{%
Cohen%
\ \protect \BOthers {.}}{%
{\protect \APACyear {1990}}%
}]{%
cohen1990}
\APACinsertmetastar {%
cohen1990}%
\begin{APACrefauthors}%
Cohen, J\BPBI D.%
, Dunbar, K.%
\BCBL {}\ \BBA {} McClelland, J\BPBI L.%
\end{APACrefauthors}%
\unskip\
\newblock
\APACrefYearMonthDay{1990}{}{}.
\newblock
{\BBOQ}\APACrefatitle {On the control of automatic processes: A parallel
  distributed processing model of the Stroop effect} {On the control of
  automatic processes: A parallel distributed processing model of the stroop
  effect}.{\BBCQ}
\newblock
\APACjournalVolNumPages{Psychological Review}{97}{}{}.
\PrintBackRefs{\CurrentBib}

\bibitem [\protect \citeauthoryear {%
Collobert%
\ \BBA {} Weston%
}{%
Collobert%
\ \BBA {} Weston%
}{%
{\protect \APACyear {2008}}%
}]{%
collobert2008unified}
\APACinsertmetastar {%
collobert2008unified}%
\begin{APACrefauthors}%
Collobert, R.%
\BCBT {}\ \BBA {} Weston, J.%
\end{APACrefauthors}%
\unskip\
\newblock
\APACrefYearMonthDay{2008}{}{}.
\newblock
{\BBOQ}\APACrefatitle {A unified architecture for natural language processing:
  Deep neural networks with multitask learning} {A unified architecture for
  natural language processing: Deep neural networks with multitask
  learning}.{\BBCQ}
\newblock
\BIn{} \APACrefbtitle {Proceedings of the 25th international conference on
  Machine learning} {Proceedings of the 25th international conference on
  machine learning}\ (\BPGS\ 160--167).
\PrintBackRefs{\CurrentBib}

\bibitem [\protect \citeauthoryear {%
Feng%
, Schwemmer%
, Gershman%
\BCBL {}\ \BBA {} Cohen%
}{%
Feng%
\ \protect \BOthers {.}}{%
{\protect \APACyear {2014}}%
}]{%
feng2014}
\APACinsertmetastar {%
feng2014}%
\begin{APACrefauthors}%
Feng, S\BPBI F.%
, Schwemmer, M.%
, Gershman, S\BPBI J.%
\BCBL {}\ \BBA {} Cohen, J\BPBI D.%
\end{APACrefauthors}%
\unskip\
\newblock
\APACrefYearMonthDay{2014}{}{}.
\newblock
{\BBOQ}\APACrefatitle {Multitasking versus multiplexing: Toward a normative
  account of limitations in the simultaneous execution of control-demanding
  behaviors} {Multitasking versus multiplexing: Toward a normative account of
  limitations in the simultaneous execution of control-demanding
  behaviors}.{\BBCQ}
\newblock
\APACjournalVolNumPages{Cognitive, Affective, \& Behavioral
  Neuroscience}{14}{1}{129--146}.
\PrintBackRefs{\CurrentBib}

\bibitem [\protect \citeauthoryear {%
Garner%
\ \BBA {} Dux%
}{%
Garner%
\ \BBA {} Dux%
}{%
{\protect \APACyear {2015}}%
}]{%
garner2015}
\APACinsertmetastar {%
garner2015}%
\begin{APACrefauthors}%
Garner, K.%
\BCBT {}\ \BBA {} Dux, P\BPBI E.%
\end{APACrefauthors}%
\unskip\
\newblock
\APACrefYearMonthDay{2015}{10}{}.
\newblock
{\BBOQ}\APACrefatitle {Training conquers multitasking costs by dividing task
  representations in the frontoparietal- subcortical system} {Training conquers
  multitasking costs by dividing task representations in the frontoparietal-
  subcortical system}.{\BBCQ}
\newblock
\APACjournalVolNumPages{Proceedings of the National Academy of Sciences}{}{}{}.
\PrintBackRefs{\CurrentBib}

\bibitem [\protect \citeauthoryear {%
Gershman%
, Horvitz%
\BCBL {}\ \BBA {} Tenenbaum%
}{%
Gershman%
\ \protect \BOthers {.}}{%
{\protect \APACyear {2015}}%
}]{%
gershman2015computational}
\APACinsertmetastar {%
gershman2015computational}%
\begin{APACrefauthors}%
Gershman, S\BPBI J.%
, Horvitz, E\BPBI J.%
\BCBL {}\ \BBA {} Tenenbaum, J\BPBI B.%
\end{APACrefauthors}%
\unskip\
\newblock
\APACrefYearMonthDay{2015}{}{}.
\newblock
{\BBOQ}\APACrefatitle {Computational rationality: A converging paradigm for
  intelligence in brains, minds, and machines} {Computational rationality: A
  converging paradigm for intelligence in brains, minds, and machines}.{\BBCQ}
\newblock
\APACjournalVolNumPages{Science}{349}{6245}{273--278}.
\PrintBackRefs{\CurrentBib}

\bibitem [\protect \citeauthoryear {%
Gigerenzer%
}{%
Gigerenzer%
}{%
{\protect \APACyear {2008}}%
}]{%
gigerenzer2008heuristics}
\APACinsertmetastar {%
gigerenzer2008heuristics}%
\begin{APACrefauthors}%
Gigerenzer, G.%
\end{APACrefauthors}%
\unskip\
\newblock
\APACrefYearMonthDay{2008}{}{}.
\newblock
{\BBOQ}\APACrefatitle {Why heuristics work} {Why heuristics work}.{\BBCQ}
\newblock
\APACjournalVolNumPages{Perspectives on psychological science}{3}{1}{20--29}.
\PrintBackRefs{\CurrentBib}

\bibitem [\protect \citeauthoryear {%
Griffiths%
, Lieder%
\BCBL {}\ \BBA {} Goodman%
}{%
Griffiths%
\ \protect \BOthers {.}}{%
{\protect \APACyear {2015}}%
}]{%
griffiths2015rational}
\APACinsertmetastar {%
griffiths2015rational}%
\begin{APACrefauthors}%
Griffiths, T\BPBI L.%
, Lieder, F.%
\BCBL {}\ \BBA {} Goodman, N\BPBI D.%
\end{APACrefauthors}%
\unskip\
\newblock
\APACrefYearMonthDay{2015}{}{}.
\newblock
{\BBOQ}\APACrefatitle {Rational use of cognitive resources: Levels of analysis
  between the computational and the algorithmic} {Rational use of cognitive
  resources: Levels of analysis between the computational and the
  algorithmic}.{\BBCQ}
\newblock
\APACjournalVolNumPages{Topics in cognitive science}{7}{2}{217--229}.
\PrintBackRefs{\CurrentBib}

\bibitem [\protect \citeauthoryear {%
Hinton%
}{%
Hinton%
}{%
{\protect \APACyear {1986}}%
}]{%
hinton_learning_1986}
\APACinsertmetastar {%
hinton_learning_1986}%
\begin{APACrefauthors}%
Hinton, G\BPBI E.%
\end{APACrefauthors}%
\unskip\
\newblock
\APACrefYearMonthDay{1986}{}{}.
\newblock
{\BBOQ}\APACrefatitle {Learning {distributed} {representations} of {concepts}}
  {Learning {distributed} {representations} of {concepts}}.{\BBCQ}
\newblock
\BIn{} \APACrefbtitle {Proceedings of the 8th {confererence} of the {Cognitive}
  {Science} {Society}} {Proceedings of the 8th {confererence} of the
  {Cognitive} {Science} {Society}}\ (\BPGS\ 1--12).
\newblock
\APACaddressPublisher{Hillsdale, NJ}{Lawrence Erlbaum Associates}.
\PrintBackRefs{\CurrentBib}

\bibitem [\protect \citeauthoryear {%
Lesnick%
, Musslick%
, Dey%
\BCBL {}\ \BBA {} Cohen%
}{%
Lesnick%
\ \protect \BOthers {.}}{%
{\protect \APACyear {2020}}%
}]{%
LesnickFormal}
\APACinsertmetastar {%
LesnickFormal}%
\begin{APACrefauthors}%
Lesnick, M.%
, Musslick, S.%
, Dey, B.%
\BCBL {}\ \BBA {} Cohen, J\BPBI D.%
\end{APACrefauthors}%
\unskip\
\newblock
\APACrefYearMonthDay{2020}{}{}.
\newblock
{\BBOQ}\APACrefatitle {A Formal Framework for Cognitive Models of Multitasking}
  {A formal framework for cognitive models of multitasking}.{\BBCQ}
\newblock

\newblock
\begin{APACrefDOI} \doi{https://doi.org/10.31234/osf.io/7yzdn} \end{APACrefDOI}
\PrintBackRefs{\CurrentBib}

\bibitem [\protect \citeauthoryear {%
Logan%
}{%
Logan%
}{%
{\protect \APACyear {1980}}%
}]{%
logan1980}
\APACinsertmetastar {%
logan1980}%
\begin{APACrefauthors}%
Logan, G\BPBI D.%
\end{APACrefauthors}%
\unskip\
\newblock
\APACrefYearMonthDay{1980}{}{}.
\newblock
{\BBOQ}\APACrefatitle {Attention and automaticity in Stroop and priming tasks:
  Theory and data} {Attention and automaticity in stroop and priming tasks:
  Theory and data}.{\BBCQ}
\newblock
\APACjournalVolNumPages{Cognitive psychology}{12}{}{523-53}.
\PrintBackRefs{\CurrentBib}

\bibitem [\protect \citeauthoryear {%
Manohar%
\ \protect \BOthers {.}}{%
Manohar%
\ \protect \BOthers {.}}{%
{\protect \APACyear {2015}}%
}]{%
manohar2015reward}
\APACinsertmetastar {%
manohar2015reward}%
\begin{APACrefauthors}%
Manohar, S\BPBI G.%
, Chong, T\BPBI T\BHBI J.%
, Apps, M\BPBI A.%
, Batla, A.%
, Stamelou, M.%
, Jarman, P\BPBI R.%
\BDBL {}Husain, M.%
\end{APACrefauthors}%
\unskip\
\newblock
\APACrefYearMonthDay{2015}{}{}.
\newblock
{\BBOQ}\APACrefatitle {Reward pays the cost of noise reduction in motor and
  cognitive control} {Reward pays the cost of noise reduction in motor and
  cognitive control}.{\BBCQ}
\newblock
\APACjournalVolNumPages{Current Biology}{25}{13}{1707--1716}.
\PrintBackRefs{\CurrentBib}

\bibitem [\protect \citeauthoryear {%
McClelland%
, McNaughton%
\BCBL {}\ \BBA {} O'Reilly%
}{%
McClelland%
\ \protect \BOthers {.}}{%
{\protect \APACyear {1995}}%
}]{%
mcclelland1995}
\APACinsertmetastar {%
mcclelland1995}%
\begin{APACrefauthors}%
McClelland, J.%
, McNaughton, B.%
\BCBL {}\ \BBA {} O'Reilly, R.%
\end{APACrefauthors}%
\unskip\
\newblock
\APACrefYearMonthDay{1995}{}{}.
\newblock
{\BBOQ}\APACrefatitle {Why There are Complementary Learning Systems in the
  Hippocampus and Neocortex: Insights from the Successes and Failures of
  Connectionist Models of Learning and Memory} {Why there are complementary
  learning systems in the hippocampus and neocortex: Insights from the
  successes and failures of connectionist models of learning and
  memory}.{\BBCQ}
\newblock
\APACjournalVolNumPages{Psychological Review}{102}{}{}.
\PrintBackRefs{\CurrentBib}

\bibitem [\protect \citeauthoryear {%
Meyer%
\ \BBA {} Kieras%
}{%
Meyer%
\ \BBA {} Kieras%
}{%
{\protect \APACyear {1997}}%
}]{%
meyer1997}
\APACinsertmetastar {%
meyer1997}%
\begin{APACrefauthors}%
Meyer, D.%
\BCBT {}\ \BBA {} Kieras, D.%
\end{APACrefauthors}%
\unskip\
\newblock
\APACrefYearMonthDay{1997}{02}{}.
\newblock
{\BBOQ}\APACrefatitle {A Computational Theory of Executive Cognitive Processes
  and Multiple-Task Performance: Part 1. Basic Mechanisms} {A computational
  theory of executive cognitive processes and multiple-task performance: Part
  1. basic mechanisms}.{\BBCQ}
\newblock
\APACjournalVolNumPages{Psychological Review}{104}{}{3-65}.
\PrintBackRefs{\CurrentBib}

\bibitem [\protect \citeauthoryear {%
Musslick%
\ \protect \BOthers {.}}{%
Musslick%
\ \protect \BOthers {.}}{%
{\protect \APACyear {2016}}%
}]{%
musslick2016}
\APACinsertmetastar {%
musslick2016}%
\begin{APACrefauthors}%
Musslick, S.%
, Dey, B.%
, {\"O}zcimder, K.%
, Mostofa, M.%
, Patwary, A.%
, Willke, T.%
\BCBL {}\ \BBA {} Cohen, J\BPBI D.%
\end{APACrefauthors}%
\unskip\
\newblock
\APACrefYearMonthDay{2016}{08}{}.
\newblock
{\BBOQ}\APACrefatitle {Controlled vs. Automatic Processing: A Graph-Theoretic
  Approach to the Analysis of Serial vs. Parallel Processing in Neural Network
  Architectures} {Controlled vs. automatic processing: A graph-theoretic
  approach to the analysis of serial vs. parallel processing in neural network
  architectures}.{\BBCQ}
\newblock
\BIn{} \APACrefbtitle {Proceedings of the 38th Annual Conference of the
  {Cognitive Science Society}} {Proceedings of the 38th annual conference of
  the {Cognitive Science Society}}\ (\BPGS\ 1547--1552).
\PrintBackRefs{\CurrentBib}

\bibitem [\protect \citeauthoryear {%
Musslick%
\ \protect \BOthers {.}}{%
Musslick%
\ \protect \BOthers {.}}{%
{\protect \APACyear {2017}}%
}]{%
Musslick_et_al_2017}
\APACinsertmetastar {%
Musslick_et_al_2017}%
\begin{APACrefauthors}%
Musslick, S.%
, Saxe, A.%
, {\"O}zcimder, K.%
, Dey, B.%
, Henselman, G.%
\BCBL {}\ \BBA {} Cohen, J\BPBI D.%
\end{APACrefauthors}%
\unskip\
\newblock
\APACrefYearMonthDay{2017}{August}{}.
\newblock
{\BBOQ}\APACrefatitle {Multitasking Capability Versus Learning Efficiency in
  Neural Network Architectures} {Multitasking capability versus learning
  efficiency in neural network architectures}.{\BBCQ}
\newblock
\BIn{} \APACrefbtitle {{Proceedings of the 39th Annual Meeting of the Cognitive
  Science Society}} {{Proceedings of the 39th Annual Meeting of the Cognitive
  Science Society}}\ (\BPG~829-834).
\PrintBackRefs{\CurrentBib}

\bibitem [\protect \citeauthoryear {%
Navon%
\ \BBA {} Gopher%
}{%
Navon%
\ \BBA {} Gopher%
}{%
{\protect \APACyear {1979}}%
}]{%
navon1979economy}
\APACinsertmetastar {%
navon1979economy}%
\begin{APACrefauthors}%
Navon, D.%
\BCBT {}\ \BBA {} Gopher, D.%
\end{APACrefauthors}%
\unskip\
\newblock
\APACrefYearMonthDay{1979}{}{}.
\newblock
{\BBOQ}\APACrefatitle {On the economy of the human-processing system.} {On the
  economy of the human-processing system.}{\BBCQ}
\newblock
\APACjournalVolNumPages{Psychological Review}{86}{3}{214}.
\PrintBackRefs{\CurrentBib}

\bibitem [\protect \citeauthoryear {%
Petri%
\ \protect \BOthers {.}}{%
Petri%
\ \protect \BOthers {.}}{%
{\protect \APACyear {2020}}%
}]{%
PetriInPrep}
\APACinsertmetastar {%
PetriInPrep}%
\begin{APACrefauthors}%
Petri, G.%
, Musslick, S.%
, Öczimder, K.%
, Dey, B.%
, Ahmed, N.%
, Willke, T.%
\BCBL {}\ \BBA {} Cohen, J\BPBI D.%
\end{APACrefauthors}%
\unskip\
\newblock
\APACrefYearMonthDay{2020}{}{}.
\newblock
{\BBOQ}\APACrefatitle {Universal limits to parallel processing capability of
  network architectures} {Universal limits to parallel processing capability of
  network architectures}.{\BBCQ}
\newblock
\begin{APACrefURL} \url{https://arxiv.org/abs/1708.03263} \end{APACrefURL}
\PrintBackRefs{\CurrentBib}

\bibitem [\protect \citeauthoryear {%
Posner%
\ \BBA {} Snyder%
}{%
Posner%
\ \BBA {} Snyder%
}{%
{\protect \APACyear {1975}}%
}]{%
posnerr}
\APACinsertmetastar {%
posnerr}%
\begin{APACrefauthors}%
Posner, M.%
\BCBT {}\ \BBA {} Snyder, C.%
\end{APACrefauthors}%
\unskip\
\newblock
\APACrefYearMonthDay{1975}{}{}.
\newblock
{\BBOQ}\APACrefatitle {“Attention and cognitive control”} {“attention and
  cognitive control”}.{\BBCQ}
\newblock
\BIn{} \APACrefbtitle {Information processing and cognition: The Loyola
  symposium} {Information processing and cognition: The loyola symposium}\
  (\BPGS\ 55--85).
\PrintBackRefs{\CurrentBib}

\bibitem [\protect \citeauthoryear {%
Russell%
\ \BBA {} Subramanian%
}{%
Russell%
\ \BBA {} Subramanian%
}{%
{\protect \APACyear {1995}}%
}]{%
russell1995provably}
\APACinsertmetastar {%
russell1995provably}%
\begin{APACrefauthors}%
Russell, S\BPBI J.%
\BCBT {}\ \BBA {} Subramanian, D.%
\end{APACrefauthors}%
\unskip\
\newblock
\APACrefYearMonthDay{1995}{}{}.
\newblock
{\BBOQ}\APACrefatitle {Provably bounded-optimal agents} {Provably
  bounded-optimal agents}.{\BBCQ}
\newblock
\APACjournalVolNumPages{Journal of Artificial Intelligence
  Research}{2}{}{575--609}.
\PrintBackRefs{\CurrentBib}

\bibitem [\protect \citeauthoryear {%
Salvucci%
\ \BBA {} Taatgen%
}{%
Salvucci%
\ \BBA {} Taatgen%
}{%
{\protect \APACyear {2008}}%
}]{%
salvucci2008threaded}
\APACinsertmetastar {%
salvucci2008threaded}%
\begin{APACrefauthors}%
Salvucci, D\BPBI D.%
\BCBT {}\ \BBA {} Taatgen, N\BPBI A.%
\end{APACrefauthors}%
\unskip\
\newblock
\APACrefYearMonthDay{2008}{}{}.
\newblock
{\BBOQ}\APACrefatitle {Threaded cognition: An integrated theory of concurrent
  multitasking.} {Threaded cognition: An integrated theory of concurrent
  multitasking.}{\BBCQ}
\newblock
\APACjournalVolNumPages{Psychological review}{115}{1}{101}.
\PrintBackRefs{\CurrentBib}

\bibitem [\protect \citeauthoryear {%
Saxe%
, McClelland%
\BCBL {}\ \BBA {} Ganguli%
}{%
Saxe%
\ \protect \BOthers {.}}{%
{\protect \APACyear {2013}}%
}]{%
saxe2013learning}
\APACinsertmetastar {%
saxe2013learning}%
\begin{APACrefauthors}%
Saxe, A\BPBI M.%
, McClelland, J\BPBI L.%
\BCBL {}\ \BBA {} Ganguli, S.%
\end{APACrefauthors}%
\unskip\
\newblock
\APACrefYearMonthDay{2013}{}{}.
\newblock
{\BBOQ}\APACrefatitle {Learning hierarchical category structure in deep neural
  networks} {Learning hierarchical category structure in deep neural
  networks}.{\BBCQ}
\newblock
\BIn{} \APACrefbtitle {Proceedings of the 35th annual meeting of the Cognitive
  Science Society} {Proceedings of the 35th annual meeting of the cognitive
  science society}\ (\BPGS\ 1271--1276).
\PrintBackRefs{\CurrentBib}

\bibitem [\protect \citeauthoryear {%
Shenhav%
, Botvinick%
\BCBL {}\ \BBA {} Cohen%
}{%
Shenhav%
\ \protect \BOthers {.}}{%
{\protect \APACyear {2013}}%
}]{%
shenhav2013}
\APACinsertmetastar {%
shenhav2013}%
\begin{APACrefauthors}%
Shenhav, A.%
, Botvinick, M.%
\BCBL {}\ \BBA {} Cohen, J\BPBI D.%
\end{APACrefauthors}%
\unskip\
\newblock
\APACrefYearMonthDay{2013}{07}{}.
\newblock
{\BBOQ}\APACrefatitle {The Expected Value of Control: An Integrative Theory of
  Anterior Cingulate Cortex Function} {The expected value of control: An
  integrative theory of anterior cingulate cortex function}.{\BBCQ}
\newblock
\APACjournalVolNumPages{Neuron}{79}{}{217-40}.
\PrintBackRefs{\CurrentBib}

\bibitem [\protect \citeauthoryear {%
Shenhav%
\ \protect \BOthers {.}}{%
Shenhav%
\ \protect \BOthers {.}}{%
{\protect \APACyear {2017}}%
}]{%
shenhav2017}
\APACinsertmetastar {%
shenhav2017}%
\begin{APACrefauthors}%
Shenhav, A.%
, Musslick, S.%
, Lieder, F.%
, Kool, W.%
, L~Griffiths, T.%
, D~Cohen, J.%
\BCBL {}\ \BBA {} Botvinick, M.%
\end{APACrefauthors}%
\unskip\
\newblock
\APACrefYearMonthDay{2017}{01}{}.
\newblock
{\BBOQ}\APACrefatitle {Toward a Rational and Mechanistic Account of Mental
  Effort} {Toward a rational and mechanistic account of mental effort}.{\BBCQ}
\newblock
\APACjournalVolNumPages{Annual Review of Neuroscience}{40}{}{}.
\PrintBackRefs{\CurrentBib}

\bibitem [\protect \citeauthoryear {%
Shiffrin%
\ \BBA {} Schneider%
}{%
Shiffrin%
\ \BBA {} Schneider%
}{%
{\protect \APACyear {1977}}%
}]{%
shiffrin1977}
\APACinsertmetastar {%
shiffrin1977}%
\begin{APACrefauthors}%
Shiffrin, R.%
\BCBT {}\ \BBA {} Schneider, W.%
\end{APACrefauthors}%
\unskip\
\newblock
\APACrefYearMonthDay{1977}{03}{}.
\newblock
{\BBOQ}\APACrefatitle {Controlled and automatic human information processing:
  {II. Perceptual} learning, automatic attending and a general theory}
  {Controlled and automatic human information processing: {II. Perceptual}
  learning, automatic attending and a general theory}.{\BBCQ}
\newblock
\APACjournalVolNumPages{Psychological Review}{84}{}{127-190}.
\PrintBackRefs{\CurrentBib}

\bibitem [\protect \citeauthoryear {%
Simon%
}{%
Simon%
}{%
{\protect \APACyear {1955}}%
}]{%
simon1955behavioral}
\APACinsertmetastar {%
simon1955behavioral}%
\begin{APACrefauthors}%
Simon, H\BPBI A.%
\end{APACrefauthors}%
\unskip\
\newblock
\APACrefYearMonthDay{1955}{}{}.
\newblock
{\BBOQ}\APACrefatitle {A behavioral model of rational choice} {A behavioral
  model of rational choice}.{\BBCQ}
\newblock
\APACjournalVolNumPages{The quarterly journal of economics}{69}{1}{99--118}.
\PrintBackRefs{\CurrentBib}

\bibitem [\protect \citeauthoryear {%
Simon%
}{%
Simon%
}{%
{\protect \APACyear {1982}}%
}]{%
simon1982models}
\APACinsertmetastar {%
simon1982models}%
\begin{APACrefauthors}%
Simon, H\BPBI A.%
\end{APACrefauthors}%
\unskip\
\newblock
\APACrefYearMonthDay{1982}{}{}.
\newblock
\APACrefbtitle {Models of bounded rationality. 1982.} {Models of bounded
  rationality. 1982.}
\newblock
\APACaddressPublisher{}{Cambridge: MIT Press}.
\PrintBackRefs{\CurrentBib}

\bibitem [\protect \citeauthoryear {%
Wickens%
}{%
Wickens%
}{%
{\protect \APACyear {1991}}%
}]{%
wickens1991processing}
\APACinsertmetastar {%
wickens1991processing}%
\begin{APACrefauthors}%
Wickens, C\BPBI D.%
\end{APACrefauthors}%
\unskip\
\newblock
\APACrefYearMonthDay{1991}{}{}.
\newblock
{\BBOQ}\APACrefatitle {Processing resources and attention} {Processing
  resources and attention}.{\BBCQ}
\newblock
\APACjournalVolNumPages{Multiple-task performance}{1991}{}{3--34}.
\PrintBackRefs{\CurrentBib}

\bibitem [\protect \citeauthoryear {%
Yin%
\ \BBA {} Knowlton%
}{%
Yin%
\ \BBA {} Knowlton%
}{%
{\protect \APACyear {2006}}%
}]{%
yin2006}
\APACinsertmetastar {%
yin2006}%
\begin{APACrefauthors}%
Yin, H.%
\BCBT {}\ \BBA {} Knowlton, B.%
\end{APACrefauthors}%
\unskip\
\newblock
\APACrefYearMonthDay{2006}{}{}.
\newblock
{\BBOQ}\APACrefatitle {The role of the basal ganglia in habit formation} {The
  role of the basal ganglia in habit formation}.{\BBCQ}
\newblock
\APACjournalVolNumPages{Nature reviews. Neuroscience}{7}{}{}.
\PrintBackRefs{\CurrentBib}

\end{thebibliography}

\end{document}